\documentstyle{aipproc}
\begin{document}
\title{Isospin Splittings of Baryons}
\author{K{\'a}lm{\'a}n Varga$^*$,
Marco Genovese$^{\dagger}$\thanks{Supported by the EU Program ERBFMBICT 
950427}, Jean-Marc Richard$^{\dagger}$ \\ and Bernard Silvestre-Brac$^{\dagger}$}
\address{$^*$Institute for Nuclear Research of the Hungarian Academy of 
Sciences,\\
Debrecen, PO Box 51, Hungary,\\
and RIKEN, Hirosawa 2-1, Wako, Saitama 35101, Japan\\
$^\dagger$Institut des Sciences Nucl\'eaires, Universit\'e Joseph Fourier--CNRS-IN2P3,\\
53, avenue des Martyrs, F--38026  Grenoble}
\maketitle
\begin{abstract} 
We discuss the isospin-breaking mass differences among baryons, with 
particular attention in the charm sector to the $\Sigma_c^{+}-\Sigma_c^0$, 
$\Sigma_c^{++}-\Sigma_c^0$, and $\Xi_c^+-\Xi_c^0$ splittings.  Simple 
potential models cannot accommodate the trend of the available data on 
charm baryons.  More precise measurements would offer the possibility 
of testing how well potential models describe the 
non-perturbative limit of QCD.
\end{abstract} 
%

A successful phenomenology of the hadron spectrum has been obtained
using non-relativistic potential models, which tentatively simulate
the low-energy limit of QCD.  Among the observables of interest,
isospin-violating mass differences
have received much attention.  In general, the $n-p$, $\Sigma^{-} -
\Sigma^{0}$, $\Sigma^{-} -
\Sigma^{+}$, $\Xi^{-} - \Xi^{0}$ splittings of the nucleon, i.e. $\Sigma$
and $\Xi$ multiplets are well reproduced, this fixing the quark-mass
difference $m_d-m_u$.  Predictions for charmed baryons can then be
supplied.  Some results concerning the $\Sigma_c$ and $\Xi_c$
multiplets are shown in Table \ref{Tab1}, together with experimental  
data
\cite{PDB}.
\begin{table}[hbct]
\caption{\label{Tab1} Predictions of different  models for charmed
baryons
electromagnetic mass splittings}
\begin{tabular}{lccc}
 Model & $\Sigma_c^{++} - \Sigma_c^{0}$& $\Sigma_c^{+} -
\Sigma_c^{0}$ & $\Xi_c^{0} - \Xi_c^{+}$\\
\hline
Experiment \protect\cite{PDB} & $0.8 \pm 0.4\;$MeV & $1.4 \pm
0.6\;$MeV& $ 6.3
\pm 2.1\;$MeV\\
 Wright \protect\cite{Wright} & $-1.4 $&$ -1.98$& 3.08 \\
Deshpande et al. \protect\cite{Desh}&$ - (3-18)$ & $- (2.5-10)$&
$4.5-12$\\
Itoh \protect\cite{Itoh} & 6.5 & 2.4 & 2.51 \\
Ono \protect\cite{Ono} & 6.1 & 2.24 & 1.77\\
Lane and Weinberg \protect\cite{LW} & $-6$ & $-4$ & 4 \\
Chan \protect\cite{Chan} & 0.4 & $-0.7$ & 3.2 \\
Lichtenberg \protect\cite{Don} & 3.4 & 0.8 & 1.1 \\
Kalman and Jakimow \protect\cite{Kalman} & $-2.7$& $-2.24$& 3.6 \\
Isgur \protect\cite{Isgur} & $-2$& $-1.8$& \\
Richard and Taxil \protect\cite{JM}
${{\displaystyle\rm I}\atop {\displaystyle\rm II}}$ &
${\displaystyle 3\atop \displaystyle -2}$ &
${\displaystyle 1\atop\displaystyle  -1}$&
${\displaystyle 0\atop \displaystyle 2}$\\
\end{tabular}
\end{table}

Some of the models include only a fraction of the possible  
contributions.
For instance, the electrostatic interaction is accounted for, but the mass
dependence of the chromomagnetic interaction is neglected when  
replacing
a $d$ quark by a $u$ quark. This is hardly justified.
As underlined, e.g., by Isgur \cite{Isgur},  these isospin splittings
arise from several canceling contributions, so that each effect  
should be
carefully computed and even small terms should be incorporated. This  
was
checked once more in the present calculation.

The most striking feature of Table \ref{Tab1} is the wide spread of
predictions.  Next comes the observation that none of the models is
compatible with the presently available data \cite{PDB}.  In  
particular the
predicted $\Xi_c^{0} - \Xi_c^{+}$ splitting tends to be smaller than
the PDG average $6.3 \pm 2.1$ MeV \cite{PDB}.  However, a preliminary
result by CLEO indicates a smaller value $\Xi_c^{0} -
\Xi_c^{+}=2.5\pm1.7\pm1.1\;$MeV \cite{CLEO}.

The $\Sigma_{c}$ multiplet is the most puzzling.  The
$\Sigma_{c}^{++}-\Sigma_{c}^{0}$ splitting is usually larger than
$\Sigma_{c}^{+}-\Sigma_{c}^{0}$, while data seemingly favour the
reverse.  In other words, 
to the extent one can draw any conclusion from the data,
most models predict an ordering of
$\Sigma_c^{++}$, $\Sigma_c^{+}$ and $\Sigma_c^{0}$ which is not 
observed.

In Table \ref{Tab3}, the electromagnetic splittings are revisited  
using the
potential models of Refs.~\cite{Bhad} (model BCN) and \cite{SBS}  
(models AL1
and AP1), supplemented by the electrostatic interaction between  
quarks.  The
quark-mass difference   $m_d- m_u$ is adjusted  to reproduce the
neutron--proton and $\Sigma^{-} - \Sigma^{+}$  mass splittings.
The magnetic  term is small \cite{VGRS}. The model-independent sum  
rules by
Franklin \cite{Franklin} are verified with good accuracy.

\begin{table}[hbct]
\caption{
\label{Tab3}
Comparison of calculated isospin-breaking splittings (in MeV) with  
experimental
results.}
\begin{tabular}{ccddd}
Splitting & Exp. [Ref] & BCN & AL1 & AP1  \\
\hline
$n-p$ &
${\displaystyle1.293318 \pm \atop\displaystyle   
0.0000009\protect\cite{PDB}}$
&
1.38 & 1.16 & 1.29  \\
$\Delta^{0}-\Delta^{++}$ & $2.7\pm0.3$\protect\cite{PDB}
&3.21 & 2.20 & 6.10  \\
$\Sigma^{-}-\Sigma^{0}$ & $4.88\pm0.08$\protect\cite{PDB}
&6.62 & 5.16 & 6.07  \\
$\Sigma^{-}-\Sigma^{+}$ & $8.09\pm 0.16$\protect\cite{PDB}
&11.96 & 8.25 & 10.57 \\
$\Sigma^{*0}-\Sigma^{*+}$ & --4 to 4\protect\cite{PDB}
& 4.10 & 1.82 & 2.65  \\
$\Sigma^{*-}-\Sigma^{*0}$ & $2.0\pm2.4$\protect\cite{PDB}
& 6.34 & 3.85 & 4.40   \\
$\Xi^{-}-\Xi^{0}$ & $6.4\pm0.6$\protect\cite{PDB}
& 10.62 &  7.12 & 9.19   \\
$\Xi^{*-}-\Xi^{*0}$ & $3.2\pm0.6$\protect\cite{PDB}
& 5.87 &  3.68 & 3.58   \\
$\Sigma_{c}^{++}-\Sigma_{c}^{0}$ & $0.8\pm0.4$\protect\cite{PDB}
& 0.12 & 1.06 & 2.91  \\
$\Sigma_{c}^{+}-\Sigma_{c}^{0}$ & $1.4\pm0.6$\protect\cite{PDB}
& --0.96 & --0.55 & 0.55 \\
$\Xi_{c}^{0}-\Xi_{c}^{+}$ & $2.5\pm1.7\pm1.1$\protect\cite{CLEO}
& 4.67  & 2.58  &  2.90   \\
${\Xi'}_{c}^{0}-{\Xi'}_{c}^{+}$ & $1.7\pm4.6$\protect\cite{CLEO}
&1.04 & 0.47 &  --0.22   \\
$\Xi_{c}^{*0}-\Xi_{c}^{*+}$ &  $6.3\pm2.6$\protect\cite{CLEO}
&0.40    & 0.44      &--0.85   \\
$\Sigma_{b}^{+}-\Sigma_{b}^{-}$ & &  --3.58 &  --3.45  &5.64  \\
$\Sigma_{b}^{0}-\Sigma_{b}^{-}$ & & --2.85 &  --1.99  &0.01  \\
$\Xi_{b}^{-}-\Xi_{b}^{0}$ & & 7.25  & 5.12  & 4.27 \\
$\Xi_{cc}^+-\Xi_{cc}^{++}$& &--1.87 & --2.70  &--5.21 \\
\end{tabular}
\end{table}

As seen from Table \ref{Tab3}, light baryons come out in good  
agreement with
the experimental data \cite{PDB}, but some problems appear
for charmed baryons (see Table \ref{Tab3}). Reasonable changes of  
light quark
masses or other parameters do not modify substantially
this situation.
In fact, while the experimental datum $\Sigma_c^{++} - \Sigma_c^{0}=
0.8 \pm 0.4$ MeV is well reproduced, one finds a small or negative
$\Sigma_c^{+} - \Sigma_c^{0}$, at
variance with the experimental value $\Sigma_c^{+} - \Sigma_c^{0} =  
1.4
\pm 0.6$ MeV. The result for $\Xi_c^{+} - \Xi_c^{0}$  MeV
is smaller than the PDG average $ 6.3 \pm 2.3 $ but closer to the PDG  
fit $4.7
\pm 2.1$ MeV or the recent CLEO result \cite{CLEO}.

This problem raises the question whether some contribution is
neglected. For example, the models used in Table \ref{Tab3} adopt
 the empirical ``1/2 rule'' $ V_{QQQ} = \sum_{i<j}V(r_{ij})/2$,
where $V(r)$ is the quark--antiquark potential. An {\sl ad-hoc}
 3-body term
\cite{Bhad,SBS} $D_3 + { A_3 / (m_1 m_2 m_3)^{b_3}}$
is then introduced to lower the mass of baryons.
As it depends on masses, this term gives a contribution
to electromagnetic mass splittings as well.
This slightly improves the description
of the electromagnetic splittings of light baryons. However,
as is evident by inspecting the three-body term, the contribution to
$\Sigma_c^{+} - \Sigma_c^{0}$ goes in the wrong direction.
We obtain typically
$\Sigma_c^{+} - \Sigma_c^{0} \simeq -0.7$ MeV, and
$\Xi_c^{+}-\Xi_c^{0} \simeq 1.5$ MeV,
with little dependence on the choice of parameters.

Another possibility which can be explored is the running of
$\alpha_s$, which leads to a reduced coupling when heavy quarks  
appear because the scale is chosen to be proportional to the masses involved
(of course problems related to the precise choice of the scale and to
the unknown $\alpha_s$ behavior at small scales emerge).
Such an effect would decrease the coupling of the spin-spin
term and does not seem to go in the right direction for changing the
order of $\Sigma_c$ multiplet.

A further contribution could come from  instantons  
\cite{Dorokhov,Bonn}.
Interesting results have been obtained on hadron spectroscopy
with models including this instanton term replacing
\cite{Dorokhov,Bonn} or supplementing \cite{SBS2} the chromomagnetic  
force. One
finds that this interaction  does not contribute
substantially to $\Sigma_{c}$ mass splittings
for it is inversely proportional to the quark masses
and is zero for a quark pair with spin $1$, thus
could not help solving the
present problem. It gives, anyway, a positive contribution, albeit
not  expected to be quite large, to $\Xi_c^{+} - \Xi_c^{0}$.

Alternative quark models with meson exchange have been recently   
revisited
by Glozman and collaborators \cite{Gloz}.  However, the extension of  
Glozman
model to heavy baryons remains problematic, notwithstanding some  
initial
attempts \cite{Gloz2}.

In conclusion, we find that, albeit a good agreement with light-quark  
baryons
(including spin excitations), even an accurate variational treatment
implementing all the realistic interactions does not permit to  
explain the
datum on   $\Sigma_c^{+} -
\Sigma_c^{0}$ on the framework of potential models based on  
one-gluon-exchange.
Also a large $\Xi_c^+-\Xi_c^0$ is not really reproducible. More  
precise data
are however required before drawing conclusions about the relevance  
of these
models for describing the confining regime of QCD., an in particular,  
the need
for new contributions, like electromagnetic penguins \cite{Penguins}.  
A more
detailed account of our investigation will be presented elsewhere  
\cite{VGRS}.

\subsection*{ Acknowledgements} Stimulating discussions with  
Fl.~Stancu and
S.~Pepin are gratefully acknowledged, as well as correspondence from  
S.C.~Timm,
J.~Franklin and T.~Goldman. We also thank A.J.~Cole for comments on  
the
manuscript.


\begin{thebibliography}{99}

\def\NCA{{Nuovo Cimento } A }
\def\NIM{{Nucl. Instrum. Methods}}
\def\NPA{{Nucl. Phys.} A }
\def\NPB{{Nucl. Phys.} B }
\def\PLB{{Phys. Lett.}  B }
\def\PRL{{Phys. Rev. Lett.} }
\def\PRD{{Phys. Rev.} D }
\def\PRC{{Phys. Rev.} C }
\def\ZPC{{Z. Phys.} C }
\def\ZPA{{Z. Phys.} A }
\def\PTP{{Progr. Th. Phys. }}
\def\LNC{{Lett. al Nuovo Cimento} }
%
\bibitem{PDB} Review of Particle Properties, R. M. Barnett {\sl et  
al.},
\PRD 54 (1996).

\bibitem{Wright} A. C. Wright, \PRD 17 (1978) 3130.

\bibitem{Desh} N. Deshpande {\sl et al.}, \PRD 15 (1977) 1885.

\bibitem{Itoh} C. Itoh {\sl et al.}, \PTP 54 (1975) 908.

\bibitem{Ono} S. Ono, \PRD 15 (1977) 3492.

\bibitem{LW} K. Lane and S. Weinberg, \PRL 37 (1976) 717.

\bibitem{Chan} L. Chan, \PRD 15 (1977) 2478.

\bibitem{Don} D.B. Lichtenberg, \PRD 16 (1977) 231.

\bibitem{Kalman} C.S. Kalman and G. Jakimow, \LNC 19 (1977) 403.

\bibitem{Isgur} N. Isgur, \PRD 21 (1980) 779.

\bibitem{JM} J.-M. Richard and P. Taxil, \ZPC 26 (1984) 421.

\bibitem{CLEO} CLEO collaboration, preprint CLEO CONF 97-29
(available at\\
http://www.lns.cornell.edu/public/CONF/1997/CONF97-29/conf\_97-29.ps); 
\\
Steven C. Timm, Contribution at this Conference.

\bibitem{Bhad} R. K. Bhaduri {\sl et al.}, \NCA 65 (1981) 376.

\bibitem{SBS} See
B. Silvestre-Brac, Few-body Systems 20 (1996) 1, and refs. therein.

\bibitem{VGRS} K. Varga, M. Genovese, J.-M. Richard and B.  
Silvestre-Brac, in
preparation.

\bibitem{Franklin} J. Franklin, \PRD 53 (1996) 564.

\bibitem{Dorokhov} A. E. Dorokhov and N. I. Kochelev, Sov. J. Nucl.  
Phys. 52
(1990) 135, A. E. Dorokhov \NPA 581 (1995) 654.

\bibitem{Bonn} See, also, W. H. Blask {\it et al.}, \ZPA 337 (1990)  
327.

\bibitem{SBS2} C. Semay and B. Silvestre-Brac, \NPA 618 (1997) 455.

\bibitem{Gloz} L.Y. Glozman and D.O. Riska, Phys. Rep. 268 (1996)  
263,
L.Y. Glozman {\it et al.}, preprint UNIGRAZ-UTP 26-06-97.

\bibitem{Gloz2} L.Y. Glozman and D. O. Riska, \NPA 603 (1996) 326.

\bibitem{Penguins} G.J. Stephenson, K. Maltman and T. Goldman, \PRD  
43 (1991)
860; L.S. Kisslinger, T. Goldman and Z. Li, hep-ph/9610312, to appear  
in  Phys.
Lett. B.

\end{thebibliography}
\end{document}